`
\documentstyle[preprint,pra,aps]{revtex}
\begin{document}
\draft
\title{Kinks in the Presence of Rapidly Varying
Perturbations}
\author{Yuri S.  Kivshar}
\address{Optical Sciences Centre
Australian National University\\ Australian Capital Territory
0200 Canberra, Australia}

\author{Niels Gr{\o}nbech-Jensen}
\address{Theoretical Division, Los Alamos
          National Laboratory\\ Los Alamos, NM 87545}

\author{Robert D. Parmentier}
\address{Dipartimento di Fisica, Universit\`{a} di
               Salerno\\ I-84081 Baronissi (SA), Italy}
\maketitle

\begin{abstract}
Dynamics of sine-Gordon kinks in the presence
of rapidly varying periodic perturbations of different
physical origins is described analytically and
numerically.  The analytical approach is based on asymptotic
expansions, and it allows to derive, in a rigorous way, an
effective nonlinear equation for the slowly varying field
component in any order of the asymptotic procedure as
expansions in the small parameter $\omega^{-1}$, $\omega$ being
the frequency of the rapidly varying ac driving force.  Three
physically important examples of such a dynamics, {\em
i.e.}, kinks driven by a direct or parametric ac force, and
kinks on rotating and oscillating background, are analysed in
detail.  It is shown that in the main order of the asymptotic
procedure the effective equation for the slowly varying field
component is {\em a renormalized sine-Gordon equation} in the
case of the direct driving force or rotating (but phase-locked
to an external ac force) background, and it is {\em the double
sine-Gordon equation} for the parametric driving force.  The
properties of the kinks described by the renormalized nonlinear
equations are analysed, and it is demonstrated analytically and
numerically which kinds of physical phenomena may be expected
in dealing with the renormalized, rather than the unrenormalized,
nonlinear dynamics. In particular, we predict several qualitatively
new effects which include, {\em e.g.}, the perturbation-induced
internal oscillations of the $2\pi-$kink in a parametrically driven
sine-Gordon model, and generation of kink motion by a pure ac
driving force on a rotating background.

\end{abstract}
\pacs{PACS numbers: 03.40.Kf, 74.40.+b, 74.50.+r, 85.25.Cp}


\section{Introduction}

As is well known, the effect of rapidly varying perturbations on the
dynamics of nonlinear systems may lead to a drastic change of the system
behavior in the sense of the {\em averaged} dynamics. In particular,
large-amplitude {\em parametric} perturbations may give
rise to {\em a stabilization} of certain types of dynamical
regimes.  A typical and famous example is a stabilization of a
reverse pendulum by parametric forced oscillations of its pivot
\cite{1} (see also the recent paper \cite{N} and references
therein), and a similar effect may be also
achieved by applying a direct ac driving force of large
amplitude \cite{2}.  Such a dynamical stabilization has its
analog in nonlinear systems with distributed parameters
supporting, in particular, novel types of kink solitons
\cite{3,4,5,6}.  However, the method which is usually used to
derive an averaged equation describing the system dynamics in
the presence of rapidly varying perturbations
is not rigorous and, as a matter of fact, it is not well
justified.  Such a method, even being very clear from the
physical point of view, uses a splitting of slow and fast
variables and subsequent averaging which is based, in fact, on
solutions of a linearized equation for fast variations where
the slowly varying coefficients are assumed to be constant
\cite{1,3,4}.  The procedure of such a linearization assumes
that the amplitude of the forced (rapidly varying) oscillations
is small, and this is certainly valid for parametrically forced
oscillations far from the parametric resonance.  For
direct ac perturbations, the forced oscillations may become
large.  To describe the dynamics in an approximate
way, the so-called ``rotating-wave approximation'' was used
without detailed mathematical justification \cite{5,6,NL}.  It
is necessary to note that the derivation of an effective
averaged equation for the slowly varying field component is an
important (and nontrivial) step of the analysis of such
systems, and in many of the cases the corresponding equation
determines the leading physical effects observed in the
presence of rapidly varying perturbations. In all the cases it
is necessary to justify the averaging procedure as well as to
estimate the influence of the higher-order contributions.
Unfortunately, the latter are beyond the usual averaging
methods.  However, as we show in the present paper, a rigorous
analysis of the effect of rapidly varying periodic
perturbations on nonlinear dynamics of ac driven damped
systems can be performed in a straightforward
way to describe the averaged dynamics with any accuracy.

The purpose of this paper is to present the
basic steps of the method mentioned above and,
selecting the sine-Gordon (SG) model as a particular
example, to describe the dynamics of kinks in
the presence of rapidly varying driving forces of very
different physical origins.  Considering the external ac driving
force to be rapidly oscillating, we apply an asymptotic
procedure based on a Fourier series where the coefficients are
assumed to be slowly varying functions on the time scale
$\omega^{-1}$, $\omega$ being the frequency of the rapidly
varying ac force which is assumed to be large.  The basic idea
to split fast and slow variables is not new, and the well-known
example is, as mentioned, a stabilization of the reverse
pendulum by oscillations of its suspension point.  However, our
analytical method to derive an effective equation for the
slowly varying field component is novel, and this method allows
us to calculate, in a self-consistent way, all the corrections
using solely asymptotic expansions rather than direct averaging
in fast oscillations. It is clear that the applicability of the
method itself is much wider than the particular examples
covered by the present paper.

The paper is organized as follows. In Sec. II we consider the
case of a direct ac driving force demonstrating the
basic steps of our analytical approach in detail. The main
result of such an analysis is the so-called ``averaged''
equation, {\em i.e.}, that describing slowly
varying system dynamics.  In the case of the direct driving
force this equation is shown to be a renormalized SG equation.
Section III presents the case of a parametric driving force
where the final equation describing the slowly varying
dynamics is the double SG equation which, as we show, may
display new features in the averaged kink dynamics, {\em e.g.},
oscillations of the excited internal mode of the kink, which is
absent in the standard SG model. The case of kink
stabilization on a rotating background by applying a rapidly
oscillating ac force is discussed in Sec. IV. There we analyse
an effect of an induced dc force on the kink motion, as well as
showing numerically that the main conclusions of our analysis may
be easily extended to cover multi-soliton dynamics.  Finally,
Sec. V concludes the paper.

\section{Direct driving force}

\subsection{Asymptotic expansions}

As the first example, we consider the case of the direct
driving force in the SG model when the system dynamics is
described by the driven damped SG equation for the field
variable $\phi(x,t)$,
\begin{equation}
\label{1}
\frac{\partial^2 \phi}{\partial t^2} - \frac{\partial^2
\phi}{\partial x^2} + \sin \phi = f - \gamma \frac{\partial
\phi}{\partial t} + \epsilon \cos (\omega t),
\end{equation}
where $f$ is a constant contribution of the driving force,
$\gamma$ is the damping coefficient, and the amplitude
$\epsilon$ of the driving force may be large (in fact, up to
the values of order of $\omega^2$).  The standard physical
application of the model (\ref{1}) is  to describe the fluxon
dynamics in long Josephson junctions (see, {\em e.g.},
\cite{McS}), so that $f$ and $\epsilon \cos (\omega t)$ are the
constant and varying components of the bias current applied to
the junction. In the subsequent analysis we consider the
direct driving force ($\sim \epsilon$) as {\em rapidly
oscillating}, {\em i.e.}, the frequency $\omega$ is assumed to
be large in comparison with the frequency gap $(=1)$ of the
linear spectrum band.  Our purpose is to derive an ``averaged''
nonlinear equation to describe the {\em slowly varying}
dynamics of the SG field.

In order to derive an averaged equation of motion,
we note first that in the case of very different time
scales the SG field  $\phi$ may be decomposed into a sum of
slowly and rapidly varying parts, {\em i.e.},
\begin{equation}
\label{2}
\phi = \Phi + \zeta.
\end{equation}
The function
$\zeta$ stands for fast oscillations around the slowly varying
envelope function $\Phi$, and the mean value of $\zeta$ during
an oscillation period is assumed to be zero so that $<\phi> =
\Phi$.  Our goal is to derive an effective equation for the
function $\Phi$. The standard way to do that is to
substitute Eq. (\ref{2}) into Eq. (\ref{1}) and to split Eq.
(\ref{1}) into two equations for slow and fast variables,
making an averaging to obtain the equation for the slowly
varying field component (see, e.g., \cite{3,4}). However, such
an approach must be properly justified for the
case when the fast oscillations {\em are not small} as it is
for the direct driving force considered here, and in a
similar problem it was proposed \cite{5} to use the so-called
``rotating wave approximation'' to find the rapidly oscillating
field component.  All these approaches, although quite
satisfactory for the first-order approximation (see, e.g.,
\cite{3,5,6}), do not allow to make the next-order expansions
to calculate higher-order corrections, and thus they
cannot be rigorously justified.  Nevertheless, as we show in
the present paper, a rigorous approach may indeed be proposed
to obtain an effective equation for the slowly varying field
component $\Phi$ {\em with any accuracy}.

The basis of our asymptotic procedure is a
Fourier series expansion with slowly varying coefficients.
We look for rapidly oscillating component $\zeta$ in the
form
\begin{equation}
\label{5} \zeta = A \cos(\omega t) + B
\sin(\omega t) + C \cos(2 \omega t) + D \sin(2 \omega t) +
\ldots \; ,
\end{equation}
where the coefficients $A$, $B,
\ldots$ are assumed to be slowly varying on the time scale
$\sim \omega^{-1}$.  Substituting the expressions (\ref{2}),
(\ref{5}) into Eq.  (\ref{1}), we note that the effective
coupling between different harmonics of the expansion
(\ref{2}), (\ref{5}) is produced by the nonlinear term $\sin
\phi$, which generates the following Fourier expansion,
\begin{eqnarray}
\sin \phi = \sin \Phi \left[ \alpha_0 + \alpha_1 \cos (\omega
t) + \alpha_2 \sin (\omega t) + \alpha_3 \cos (2\omega t) +
\alpha_4 \sin (2\omega t) + \ldots \right] \nonumber \\
+  \cos \Phi \left[ \beta_0 + \beta_1 \cos (\omega t) + \beta_2
\sin (\omega t) + \beta_3 \cos (2\omega t) + \beta_4 \sin
(2\omega t) + \ldots  \right],
\end{eqnarray}
where
\begin{eqnarray}
\alpha_0 = J_0(A) \left( 1 -\frac{1}{4} B^2\right) +
\frac{1}{4} B^2 J_2 (A) + \ldots , \;\;\; \alpha_1 = - C J_1
(A) + \ldots , \nonumber \\
\alpha_2 = -D J_1(A) + \ldots , \;\;\; \alpha_3 = \frac{1}{4}
B^2 J_0(A) + \ldots, \;\;\; \alpha_4 = - BJ_1(A) + \ldots ,
\end{eqnarray}
\begin{eqnarray}
\beta_0 = - CJ_2(A) + \ldots, \;\;\; \beta_1 = 2J_1(A) +
\ldots, \;\;\; \beta_2 = B J_0(A) + BC J_1(A) + \ldots,
\nonumber \\
\beta_3 = C J_0(A) + \ldots, \;\;\; \beta_4 = D J_0(A) +
\ldots,
\end{eqnarray}
and $J_0$, $J_1$, {\em etc.}, are Bessel functions.
Collecting now the coefficients
in front of the different harmonics, we obtain an infinite set
of coupled nonlinear equations,
\begin{eqnarray}
\label{4}
\frac{\partial^2 \Phi}{\partial t^2} -
\frac{\partial^2 \Phi}{\partial x^2}  + \sin \Phi \left[ J_0(A)
\left(1 - \frac{1}{4} B^2\right) + \frac{1}{4} B^2 J_2(A) +
\ldots \right]
\nonumber\\
+ \cos \Phi \left[ - C J_2(A) +
\ldots \right] = f - \gamma \frac{\partial \Phi}{\partial t},
\end{eqnarray}
\begin{eqnarray}
\label{6}
\left(- \omega^2 A +
2 \omega \frac{\partial B}{\partial t} + \frac{\partial ^2
A}{\partial t^2}\right) - \frac{\partial^2 A}{\partial x^2} +
\cos \Phi [ 2 J_1(A) + \ldots ]
\nonumber\\
+ \sin \Phi [ - C
J_1(A) + \ldots ] + \gamma \left( \frac{\partial A}{\partial t}
+ \omega B\right) = \epsilon,
\end{eqnarray}
\begin{eqnarray}
\label{7}
\left( -\omega^2 B - 2 \omega \frac{\partial A}{\partial t} +
\frac{\partial^2 B}{\partial t^2} \right) - \frac{\partial^2
B}{\partial x^2} + \cos \Phi [ B J_0(A) + B C J_1(A) + \ldots ]
\nonumber\\ + \sin \Phi [ - D J_1(A) + \ldots ] + \gamma \left(
\frac{\partial B}{\partial t} - \omega A \right) = 0,
\end{eqnarray}
\begin{eqnarray}
\label{8}
\left( - 4\omega^2 C + 4\omega \frac{\partial D}{\partial t}
+ \frac{\partial^2 C}{\partial t^2} \right) -
\frac{\partial^2 C}{\partial x^2} + \cos \Phi [ C J_0(A) +
\ldots ] \nonumber \\ +\sin \Phi \left[ \frac{1}{4} B^2 J_0(A)
+ \ldots \right] + \gamma \left(\frac{\partial C}{\partial t} +
2\omega D\right) = 0,
\end{eqnarray}
\begin{eqnarray}
\label{9} \left(- 4\omega^2 D - 4\omega
\frac{\partial C}{\partial t} + \frac{\partial^2 D}{\partial
t^2}\right) - \frac{\partial^2 D}{\partial x^2} + \cos
\Phi [ D J_0(A) + \ldots ] \nonumber \\ + \sin \Phi [ - B
J_1(A) + \ldots ] + \gamma \left( \frac{\partial D}{\partial t}
- 2\omega C \right) = 0,
\end{eqnarray}
and similar equations for the
coefficients in front of the higher-order harmonics. To proceed
further, we note that different terms in Eqs.  (\ref{6}) to
(\ref{9}) are not equivalent provided $\omega$ is a large
parameter. Indeed, if we assume the amplitude $\epsilon$
large as well (otherwise, the dynamics of the system is rather
trivial because the effect of small-amplitude but rapidly
oscillating force is negligible), let us say up to the order of
$\omega^2$, the large term $-\omega^2 A$ in Eq. (\ref{6}) may
be compensated only by the term $\epsilon$ from the right-hand
side of Eq. (\ref{6}). Thus, assuming $\epsilon \sim \omega^2$
we find the first term of the asymptotic expansion $A \approx -
\epsilon/\omega^2$. On the other hand, the right-hand side of
Eq. (\ref{7}) is zero, and the large term $-\omega^2 B$ may be
compensated only by a contribution from the other terms $\sim
A$, thus giving the first term of the expansion for the
coefficient $B$, {\em viz.}, $B \approx \gamma \epsilon/\omega^3$.
Such a simple reasoning may be effectively applied to other
coefficients as well as to other corrections of the asymptotic
expansion. As a matter of fact, to generalize and simplify the
procedure of calculation of the expansion coefficients, we
look for the coefficients $A, B, \ldots$ in the form of the
power series in the small parameter $\omega^{-1}$ as
follows \begin{equation}
\label{11}
A=a_{1}+\frac{a_{2}}{\omega^2}+ \ldots, \;\;
B=\frac{b_{1}}{\omega}+\frac{b_{2}}{\omega^3} + \ldots, \;\; C
= \frac{c_{1}}{\omega^4} + \frac{c_2}{\omega^6} + \ldots, \;\;
D=\frac{d_{1}}{\omega^3}+ \frac{d_2}{\omega^5} + \ldots \;.
\end{equation} Substituting Eq. (\ref{11}) into Eqs. (\ref{6})
to (\ref{9}) and equating the terms of the same orders in the
small parameter $\omega^{-1}$, we find \begin{equation}
\label{12} a_{1}= - \frac{\epsilon}{\omega^2} \equiv - \delta,
\end{equation} \begin{equation} \label{13} a_{2}= \gamma b_1 +
2 \cos \Phi \, J_1(a_1), \end{equation} \begin{equation}
\label{14}
b_{1}= -  \gamma a_1,
\end{equation}
\begin{equation}
\label{15}
b_{2}= - 2 \frac{\partial a_2}{\partial t} - \gamma a_2  + b_1
\cos \Phi [ J_0 (a_1) + 2 J_2(a_1)],
\end{equation}
\begin{equation}
\label{16}
c_{1}=  \frac{1}{4} \left[ 4
\frac{\partial d_1}{\partial t} + \frac{1}{4} b_1^2 J_0(a_1)
\sin \Phi + \frac{1}{2} b_1^2 J_2(a_1) + 2\gamma d_1 \right],
\end{equation}
\begin{equation}
\label{16a} d_{1}= -
\frac{1}{4} b_1 J_1(a_1) \sin \Phi,
\end{equation}
and so on. In Eq. (\ref{12}) the parameter
$\delta = \epsilon/\omega^2$ is assumed to be of order of
${\cal O}(1)$, but all the results are valid also for the case
$\delta \ll 1$.  The expansions (\ref{11}) allow to find the
coefficients $A, B, \ldots$ in each order of $\omega^{-1}$, and
all the corrections are determined by {\em algebraic}
relations rather than additional differential equations.  For
example, $a_{2}$ is determined by Eq.  (\ref{13}) through
$b_{1}$ which, in its turn, may be found from Eq. (\ref{14}) as
a function of $a_{1}$, {\em i.e.}, through the slowly varying part
$\Phi$, and so on. This statement is valid for all coefficients
of the asymptotic expansion: The coefficients are found through
{\em algebraic} relations involving lower-order terms of the
asymptotic expansions and their derivatives, and one does not
need to find solutions of additional differential equations.

\subsection{Renormalized equation}

Applying the expansions (\ref{11}) to
Eq. (\ref{4}), we may find the equation for the slowly varying
field component $\Phi$ with any accuracy in the small parameter
$\omega^{-1}$, e.g.,
\begin{eqnarray}
\label{17}
\frac{\partial^2 \Phi}{\partial t^2} -
\frac{\partial^2 \Phi}{\partial x^2} + \sin \Phi \left[
J_0(a_1) + \frac{1}{\omega^2} \left( - \frac{1}{4} b_1^2
J_0(a_1) - a_2 J_1(a_1) + \frac{1}{4} b_1^2 J_2(a_1) \right) +
\ldots \right] \nonumber\\ + \cos \Phi \left[
-\frac{c_1}{\omega^4} J_2(a_1) + \ldots \right] = f - \gamma
\frac{\partial \Phi}{\partial t}.
\end{eqnarray}
Thus, from the asymptotic procedure described above it is
quite obvious how to calculate the corrections of the first,
second, and subsequent orders and to find the averaged equation
with any required accuracy.

In the first-order approximation in $\omega^{-1}$ only the term
$ J_0(a_1) \sin \Phi$ contributes, so that Eq. (\ref{17})
yields \begin{equation} \label{18} \frac{\partial^2
\Phi}{\partial t^2} - \frac{\partial^2 \Phi}{\partial x^2} +
J_0 \left(\frac{\epsilon}{\omega^2}\right) \sin \Phi = f -
\gamma \frac{\partial \Phi}{\partial t}.  \end{equation}

Equation (\ref{18}) takes
into account an effective contribution of the rapidly
varying force to the average nonlinear dynamics
and this contribution might become large for $\delta = {\cal O}(1)$,
i.e. when $\epsilon \sim \omega^2$.
Thus, the dynamics of the SG model
with a rapidly varying direct driving force may be described
by a {\em renormalized}  SG equation
(\ref{18}) up to the terms of order of $\epsilon/\omega^2$.

The results obtained above may immediately be
applied to describe the renormalized dynamics of kinks in the
presence of the rapidly varying ac force. In fact, Eq.
(\ref{18}) is the dc driven damped SG equation with a {\em
renormalized} coefficient in front of the term $\sim \sin
\Phi$. This simply means that we can apply all the results
known for the standard SG equation (see, e.g., \cite{McS,KM})
making only a {\em renormalization} of the kink's width. For
example, the kink solution of Eq.  (\ref{18}) at $\gamma = f =
0$ has the form
\begin{equation}
\label{A1} \Phi (x,t) = 4
\sigma \tan^{-1} \exp \left[ \frac{x - Vt}{l_0 \sqrt{1-V^2}}
\right],
\end{equation}
where $\sigma = \pm 1$ is the kink's
polarity and $l_0 = [J_0(\epsilon/\omega^2)]^{-1/2}$ is the
kink's width at rest. The motion of the kink in the presence of
small dc force $f$ and damping $(\sim \gamma)$ is characterized
by the steady-state velocity
\begin{equation}
\label{A2} V_{*}
= - \frac{\sigma}{\sqrt{1 + g^2}}, \;\;\;\;\; g \equiv
\left(\frac{4\gamma}{\pi f} \right)
J_0\left(\frac{\epsilon}{\omega^2} \right).
\end{equation}
In the theory of long Josephson junctions the kink's velocity is
connected with the voltage across the junction, $<\phi_t>$,
where $<\ldots>$ stands for the averaging in time, so that the
result (\ref{A2}) for the steady-state kink velocity gives the
so-called zero-field steps in the current-voltage (IV)
characteristics of a long junction.  As follows from Eq.
(\ref{A2}), the renormalization of the parameter $g$ leads
to a change of the kink's velocity $V_{*}(f)$ and this,
therefore, changes the slopes of the voltage steps by the
effect of the ac driving force.

\section{Parametric driving force}

Let us consider now a parametric driving force applied to the
SG system, with the main purpose to demonstrate that such a case
is {\em very different} from that analysed above.  The qualitative
difference between the effects produced by direct and
parametric (rapidly oscillating) forces is the following: A
sufficient change of the system ``averaged'' dynamics due to a
rapidly oscillating direct force may be observed for
amplitudes $\epsilon \sim \omega^2$ whereas in the case of a
parametric force, similar effects may be already observed
for {\em smaller} amplitude, {\em i.e.}, in fact for $\epsilon \sim
\omega$.  To prove this statement and to show how our
asymptotic method works for the case of the parametric force,
we consider the parametrically perturbed SG equation in the
form
\begin{equation}
\label{1s}
\frac{\partial^2
\phi}{\partial t^2} - \frac{\partial^2 \phi}{\partial x^2} +
\sin \phi = f - \gamma \frac{\partial \phi}{\partial t} +
\epsilon \sin \phi \cos (\omega t),
\end{equation}
where $f$ and $\gamma$ have the same sense as above, but this
time $\epsilon$ is the amplitude of the parametric force.
Various applications of the model (\ref{1s}) were discussed in
the review paper \cite{KM} (see also Refs. \cite{PSG}).

We assume that the parametric force is rapidly oscillating,
{\em i.e.}, the frequency $\omega$ is large.  As above,
we look for a solution of Eq.  (\ref{1s}) in the form of
asymptotic expansion
\begin{equation} \label{B1} \phi = \Phi +
A \cos(\omega t) + B \sin (\omega t) + C \cos (2\omega t) + D
\sin (2\omega t) + \ldots, \end{equation} where the functions
$\Phi$, $A$, $B, \ldots$ are assumed to be slowly varying on
the time scale $\sim \omega^{-1}$. The function $\Phi$ in Eq.
(\ref{B1}) determines, in fact, the evolution of the averaged
field component because $<\phi> = \Phi$, where the
brackets $<\ldots>$ stand for the averaging in fast
oscillations.  Substituting the expression (\ref{B1}) into Eq.
(\ref{1s}) and collecting, as in the case of the direct driving
force,  all the coefficients in front of the different
harmonics, we again obtain an infinite set of coupled nonlinear
equations.  The subsequent (and very important) step of such an
analysis is to find the form of the asymptotic expansions for
the coefficients $A$, $B, \ldots$. In the present case it is
easy to check that the expansions (\ref{11}) do not give a
closed asymptotic procedure, and in the case $\epsilon \sim
\omega^{2}$ the driving force from Eq.  (\ref{1s})
contributes to all the harmonics,
so that contributions of the other harmonics become large as
well.  Comparing, as in the previous case, different terms of
the equations for the coefficients $A, B, \ldots$, we may
easily check that the asymptotic procedure may be effectively
formulated for {\em smaller} (but not small) amplitudes, {\em i.e.},
when $\epsilon \sim \omega$, and, as above, it gives all the
corrections to the averaged nonlinear dynamics in a rigorous
way. Thus, we take the asymptotic expansions in the form
\begin{equation}
\label{11s} A=\frac{a_{1}}{\omega^2}+\frac{a_{2}}{\omega^4}+
\ldots, \;\; B=\frac{b_{1}}{\omega^3}+\frac{b_{2}}{\omega^5} +
\ldots, \;\; C = \frac{c_{1}}{\omega^4} + \ldots, \;\;
D=\frac{d_{1}}{\omega^5}+ \ldots \;.
\end{equation}
Using the power series (\ref{11s}), we obtain
the ``averaged'' equation in the form
\begin{equation}
\label{12s}
\frac{\partial^2 \Phi}{\partial t^2} -
\frac{\partial^2 \Phi}{\partial x^2} + \left( 1 - \frac{1}{4}
\frac{a_1^2}{\omega^4} + \ldots \right) \sin \Phi = f - \gamma
\frac{\partial \Phi}{\partial t} + \frac{\epsilon}{2} \cos \Phi
\left( \frac{a_1}{\omega^2} + \frac{a_2}{\omega^4} + \ldots
\right), \end{equation}
The expansions (\ref{11s}) allow to find the coefficients
of the asymptotic expansions in
each order in the small parameter $\omega^{-1}$, and all the
corrections are determined, as above, by {\em algebraic}
relations.

In the first-order approximation only the term $\sim
\epsilon a_1$ contributes to Eq. (\ref{12s}). From the
asymptotic expansions it follows that
\begin{equation}
\label{13s}
a_1 = - \epsilon \sin \Phi,
\end{equation}
and Eq. (\ref{12s}) yields
\begin{equation}
\label{14s} \frac{\partial^2 \Phi}{\partial t^2} -
\frac{\partial^2 \Phi}{\partial x^2} + \left( 1 + \frac{1}{2}
\Delta^2 \cos \Phi \right) \sin \Phi = f - \gamma
\frac{\partial \Phi}{\partial t},
\end{equation}
where $\Delta \equiv \epsilon /\omega$.
Equation (\ref{14s}) takes into account an effective
contribution of the rapidly varying parametric force to the
slowly varying nonlinear dynamics {\em in the lowest order}, and
all the corrections coming from the approximation of the next
order are proportional to the small parameter $\sim
\omega^{-2}$.  However, even the lowest-order contribution
might become large for $\Delta = {\cal O}(1)$, i.e. when
$\epsilon \sim \omega$.

Thus, the ``averaged'' dynamics of the SG model
with a rapidly varying parametric force is described
by the double SG equation (\ref{14s}). As a matter of fact,
the double SG equation is rather well studied (see, e.g., Refs.
\cite{Con,Sod} and references therein) and properties of its
kink solutions are known as well. In particular, the kink
solution of Eq. (\ref{14s}) at $f = \gamma = 0$ may be written
in the form \cite{Con}
\begin{equation}
\label{R1}
\Phi(x,t) =
2 \tan^{-1} \left[ \frac{1}{\sqrt{1 + \Delta^2/2}} {\rm cosech}
\left( \sqrt{1 + \frac{\Delta^2}{2}} \frac{x - Vt}{\sqrt{1 -
V^2}} \right) \right],
\end{equation}
and this solution may be
treated as two coupled $\pi$-kinks. In Fig. 1 we show the
results of numerical simulations of the parametrically driven
SG system, Eq. (23). In all the cases analysed in the present paper we
have integrated the driven damped SG equation on a spatial
interval of length $L$, with periodic boundary conditions. As is seen
from Fig. 1, the sech-type shape of the $2\pi$ kink
corresponding to the standard (unperturbed) SG system is
modified, and the function $\phi_x$ displays a two-peaked
profile which, as a matter of fact, is one of the main features
of the kink solution (\ref{R1}). Increasing $\Delta^2$ one may
observe, in accordance with Eq. (\ref{R1}), that the function
$\phi$ has an evident shape of two $\pi$-kinks separated by
a distance $\sim \Delta^2$.

As has been shown in Ref. \cite{3}, $\pi-$kinks
themselves may exist in the
parametrically driven SG chain provided the condition $\Delta^2
>2$ is satisfied. This condition simply means that the
effective ``averaged'' potential for the slowly varying field
component $\Phi$ exhibits a local minimum at $\Phi = \pi$ so
that this stationary state becomes stable.

The appearence of new features in the slowly varying
(``averaged'') system dynamics of the SG system for $\Delta^2
>2$ is similar to the phenomenon of the parametric
stabilization of the reverse pendulum in the well known Kapitza
problem \cite{1,N}.  However, in the problem under
consideration some novel features in the nonlinear dynamics of
the parametrically driven SG system may be really observed for
{\em any value of the effective parameter} $\Delta^2$.  Indeed,
as is known from the theory of the double SG equation
\cite{Sod}, at any value of $\Delta^2$ the kink (\ref{R1})
possesses the so-called internal (``shape'') mode which
describes variations of the kink's width. This internal mode is
absent for the standard SG kink, and the mode frequency
$\Omega^2_{sol}$ splits at any $\Delta^2 \neq 0$ from the gap
frequency of the linear spectrum.
For $\Delta^2 >2$ the kink's internal mode may
be described as relative oscillations of the $\pi-$ kinks of
which the $2\pi-$kink consists.  However, this mode does,
in fact, exist at any value of the effective parameter
$\Delta^2$, and it may be observed as periodic variations of
the kink's shape.

We have measured numerically the frequency of the shape
oscillations of the $2\pi$-kink (\ref{R1}) directly solving Eq.
(\ref{1s}) and also using the averaged equation (\ref{14s}).
The numerical results are shown in Fig. 2 for selected
values of the external frequency, $\omega = 50$ and $\omega =
100$. For relatively small $\Delta^2$ ({\em i.e.}, $\epsilon$ in Fig.
2), when higher-order corrections to Eq. (\ref{14s}) are
negligible, a perfect agreement between the results for the
parametrically driven SG model (\ref{1s}) and those for the
averaged equation (\ref{14s}) are clearly observed,
justifying the validity of our asymptotic procedure.

\section{Kinks on rotating and oscillating backgrounds}

As was mentioned in Ref. \cite{6}, the other physically
important case when a rapidly varying ac force may change
drastically the
kink dynamics is the case of a rotating and oscillating
background. We should note, however, that if one considers
relatively small system's length $L$, even a
relatively weak driving force may lead to complicated dynamics
involving coexisting states of bunched kinks and nontrivial
background states \cite{Rot}. These latter effects are probably
caused by the influence of nonzero boundary conditions which
may ``help to lock'' kink-like states on rotating backgrounds.
Here we are interested in the dynamics of the long SG
systems (i.e. the kink's length is much smaller than the
system length $L$) when the high-frequency force phase-locks
the SG field in an oscillating and rotating state and thereby
creates a mechanizm (an effective gravitation field) for
supporting kink solitons. However, we should note that the
theory presented below can not be applied to formally infinite
SG system because for the case of the continum linear spectrum
the applied ac force may create resonances making the system
dynamics much more complicated and even chaotic. In fact, we
need finite-width (but of large $L$) systems in order to avoid
linear resonances if the frequency of th external ac force is
selected in a gap between the nearest eigenfrequencies.

To describe the effect of the kink phase-locking on a rotating
background analytically in a rigorous way, we consider the
perturbed SG equation (\ref{1}) assuming $f>1$, in which case
the ground state of the SG chain is not stable and the chain
rotates with the frequency $\Omega$ so that $\phi \approx
\Omega t$.  Applying the high-frequency ac force $\sim
\epsilon$ we are interested in the slowly varying phase-locked
system dynamics on such a rotating background.  Accordingly, we
look for a solution of Eq. (\ref{1}) in the form
\begin{equation} \label{c2} \phi = \Phi + \Omega t + \xi,
\end{equation} where $\xi$ is the rapidly varying part
oscillating with the large frequency $\omega$ of the external
ac force, $\Phi$ is the slowly varying (long time scale) part,
and $\Omega$ is the average frequency of rotation for the
background field, which we assume to be phase-locked to the
external ac field, i.e.  $\Omega = \pm k \omega$, $k$ being
integer. Looking for the rapidly oscillating part $\xi$ in the
form of a Fourier series with slowly varying coefficients,
\begin{equation} \xi = A \cos(\omega t) + B \sin (\omega t) +
\ldots, \end{equation} we obtain the following equations for
the averaged field component $\Phi$ and the expansion
coefficients $A, B, \ldots$, \begin{equation} \label{c3}
\frac{\partial^2 \Phi}{\partial t^2} - \frac{\partial^2
\Phi}{\partial x^2} + J_k(A) \sin \Phi = f - \gamma k \omega -
\gamma \frac{\partial \Phi}{\partial t}, \end{equation}
\begin{equation} \label{c4} \left(- \omega^2A - 2\omega
\frac{\partial B}{\partial t} + \frac{\partial^2 A}{\partial
t^2} \right) - \frac{\partial^2 A}{\partial x^2} + [J_{k+1}(A)
- J_{k-1}(A)] \cos \Phi + \gamma \left( \frac{\partial
A}{\partial t} + \omega B \right) = \epsilon, \end{equation}
\begin{equation} \label{c5} \left(- \omega^2B + 2\omega
\frac{\partial A}{\partial t} + \frac{\partial^2 B}{\partial
x^2}\right) - \frac{\partial^2 B}{\partial x^2} - [J_{k+1}(A) +
J_{k-1}(A)] \sin \Phi + \gamma \left( \frac{\partial
B}{\partial t} - \omega A\right) = 0, \end{equation} and so on.
Unlike the case considered in Sec.  II, in the present problem
there are two rapidly oscillating contributions with the
frequencies $\omega$ and $k \omega$, so that the final
equations (\ref{c3}) to (\ref{c5}) for the slowly varying
coefficients differ from the corresponding equations (\ref{4})
to (\ref{7}) of Sec. II. To take into account the second term
in the right-hand side of Eq.  (\ref{c3}) in a self-consistent
way, we also assume the dissipation to be rather small, $\gamma
\omega \sim 1$, which is a typical case for Josephson
junctions.

Making now asymptotic expansions similar to the case of the
direct ac force considered above, {\em i.e.},
\begin{equation}
\label{c6} A = a_1 + \frac{a_2}{\omega^2} + \ldots, \;\; B =
\frac{b_1}{\omega} + \ldots, \end{equation} we find the
following relations [{\em cf.} Eqs. (\ref{12}), (\ref{14})]
\begin{equation} \label{c7} a_1 = - \frac{\epsilon}{\omega^2}
\equiv - \delta, \;\;\; b_1 = - \gamma a_1, \end{equation}
which allow to obtain the effective equation for the slowly
varying system dynamics by just combining Eqs. (\ref{c7}), (\ref{c6})
and (\ref{c3}). The final equation is just Eq. (\ref{c3}) with $A= -
\epsilon/\omega^2$, which describes the kink dynamics on the
background rotating with the frequency $\Omega = \pm k \omega$.
It is important to note that the resulting (effective) dc force
in the ``averaged'' nonlinear equation (\ref{c3}) is
represented by the term $f - \gamma k \omega$ but not $f$
itself, {\em i.e.}, the kink on the rotating and oscillating
background may move even in the absence of the constant
contribution to the bias current, $f=0$. Figure 3 shows the
results of the numerical calculation of the first Shapiro step ($k =
1$) of a long Josephson junction described by the model (\ref{1})
with the parameters $\gamma =0.1$, $\epsilon = 12.5$, and
$\Omega = \omega = 2.5$.  As is clearly seen from that figure, the
steps cross the zero current axis, displaying the property mentioned
above; constant-voltage zero-crossing steps are of considerable
practical interest for voltage-standard applications of Josephson
junctions (see, {\em e.g.}, \cite{FMP} and references therein).  If
we select $f=0.25$, the effective force acting on the kink vanishes,
and the kink is observed at rest (see Fig. 4).  This result is in
excellent agreement with the averaged SG equation (\ref{c3}), where
the effective force is found to be $f-\gamma k \omega$. Increasing
the value of the bias current $f$, we create an effective force
acting on the kink and it moves to the left (see Fig. 5). It is very
interesting to note that the kink motion is even possible in the
absence of the constant component of the bias current, {\em i.e.}, at
$f=0$. Figure 6 shows this case, when the effective force $-\gamma k
\omega$ generates the motion of the kink in the direction opposite to
that shown in Fig. 5.

As a matter of fact, our approach and the resulting equation
for the slowly varying field component $\Phi$ do not specify
exactly the type of nonlinear solutions we deal with. This
means that the consideration described above may be effectively
applied to other problems, much more general than a single-kink
propagation. One of the important generalizations
of this approach is to treat
multi-soliton (multi-kink) problems. In particular, for the
nonlinear dynamics on oscillating and rotating background the
effects similar to those described for a single kink may be
also observed for the case of two,
three, and more kinks. In particular, Figures
7 and 8 present results of our numerical simulations of the
same effects as for the single kink, but for the case of two
kinks in the directly driven SG model (\ref{1}). As may be
noted from these figures, the large-amplitude direct ac
force generates some radiation,
especially for the cases where the kinks are observed at
rest, but still kinks exist as localized and well defined
objects.

\section{Conclusions}

In conclusion, we have analysed the dynamics of SG kinks
in the presence of rapidly varying periodic perturbations of
different physical nature. We have proposed a rigorous
analytical approach to derive the ``averaged'' equation for the
slowly varying field component, and we have demonstrated
that in the main order of the asymptotic procedure the
effective equation is a renormalized SG equation in the case
of the direct driving force or rotating (and phase-locked to
the external ac driving force) background, and it is a double
SG equation for the parametric driving force.  However, the
method itself does allow to find in a rigorous way the
effective equation for the slowly varying field component in
any order of the asymptotic expansion in the parameter
$\omega^{-1}$, $\omega$ being the frequency of the
rapidly varying perturbations which has been assumed to be
large.

Our main purpose was to show which kinds of qualitatively new
physical effects may be expected in dealing with the
renormalized nonlinear dynamics instead of unrenormalized one.
In particular, we have predicted that the parametric driving
force may support oscillations of the kink's shape (absent in
the SG model)  which may be viewed as creation of a shape mode
of the $2\pi-$kink characterized by the internal frequency
$\Omega_{sol}$.  For the problem of the kink propagation on
rotating and oscillating background, we have shown that a
periodic ac force produces a drift in the kink motion, which may be
understood as an effect described by an effective dc force to
the kink motion in the fremework of an ``averaged'' nonlinear
dynamics.

One of the main conclusions of our analysis and
numerical simulations, {\em i.e.}, that the ``averaged''
nonlinear dynamics is drastically modified by rapidly varying
(direct or parametric) driving force but still may be
effectively described by renormalized nonlinear equations, is
rather general and applicable to many other nonlinear models
supporting various kinds of solitons.

\acknowledgments

YuK acknowledges hospitality of the Dipartimento di
Fisica dell'Universit\`{a} di Salerno during his short stay
there, and also useful discussions with A.  S\'anchez
(Madrid) and K.H. Spatschek (D\"usseldorf).  His work was
partially supported by the Australian Photonics Cooperative
Research Center (APCRC). Parts of this works were performed
under auspices of the US Department of Energy.
RDP acknowledges financial support from the EC under contract no.
SC1-CT91-0760 (TSTS) of the ``Science'' program, from MURST
(Italy), and from the Progetto Fianalizzato ``Tecnologie
Superconduttive e Criogeniche'' del CNR (Italy).

\begin{figure}
\caption{The steady-state profile of the $\phi$ (a) and
$\phi_x$ (b) fields as a function of space in the case of a
parametric driving force. The parameters are $\gamma = 0.2$, $L
= 10$, $\omega = 100$, and $\epsilon = 200$.}
\end{figure}

\begin{figure}
\caption{The frequency of the internal
oscillations of the $2\pi$-kink, $\Omega_{sol}$, as a function
of the amplitude of the ac parametric force $\epsilon$ for two
values of the external force frequency, $\omega = 50$ (squares
and crosses) and $\omega = 100$ (diamonds and plusses).
The results presented by diamonds and squares are obtained
using the effective double SG equation (\protect\ref{14s}) whereas the
plusses and crosses are the result of direct integration of the
parametrically driven SG equation (\protect\ref{1s}). Note that the
agreement between the parametrically driven model (\protect\ref{1s})
and the effective (``averaged'') model (\protect\ref{14s}) is better
for smaller $\Delta = \epsilon/\omega$, when corrections to Eq.
(\protect\ref{14s}) from the higher-order terms are negligible.}
\end{figure}

\begin{figure}
\caption{The normalized IV curves (for the first Shapiro step)
characterizing the SG dynamics with the parameters $\gamma
=0.1$, $L = 16$, $\epsilon = 12.5$, and $\Omega = 2.5$; and
with periodic boundary conditions. Shown are the
zero-field steps at $n= 0, 1, 2, 3, 4$, and $5$, $n$ being the
number of kinks in the system. Note that the IV curves
clearly cross the zero current axis and the steps are slightly
asymmetric around the voltage $<\phi_t> = 2.5$; the latter
effect is caused by the background oscillations and relatively
small velocity.}
\end{figure}

\begin{figure}
\caption{The steady-state profile of the $\phi$ (a) and
$\phi_{x}$ (b) fields as a function of space over 10 periods
of the external driving force. Parameters are $L=16$, $\gamma
=0.1$, $\epsilon = 12.5$, $\omega =2.5$, and $f =0.25$. The
value of $f$ is selected at the center of the first Shapiro
step and, therefore, the kink does not move as follows from the
theory because the effective force acting on a kink,
$f-\gamma k\omega$, is zero.}
\end{figure}

\begin{figure}
\caption{The same as in Fig. 4 but at $f=0.5$.}
\end{figure}

\begin{figure}
\caption{The same as in Fig. 4 but at $f=0$. The kink is moving
due to the uncompensated contribution of dissipative losses
$ -\gamma k \omega $.}
\end{figure}

\begin{figure}
\caption{The same as in Fig. 4 but for the case of two kinks
introduced by a change of the boundary conditions.}
\end{figure}

\begin{figure}
\caption{The same as in Fig. 5 but for the case of two kinks.}
\end{figure}
\end{document}